\documentclass{aa}  

\usepackage{graphicx}
\usepackage[varg]{txfonts}
\usepackage{hyperref}
\newcommand{\teff}{\ensuremath{T_{\mathrm{eff}}}}
\newcommand{\logg}{\ensuremath{\log g}}
\newcommand{\vmicro}{\ensuremath{\xi_\mathrm{micro}}}
\newcommand{\feh}{\ensuremath{\left[\mathrm{Fe}/\mathrm{H}\right]}}
\newcommand{\mh}{\ensuremath{\left[\mathrm{M}/\mathrm{H}\right]}}

\newcommand{\nafe}{\ensuremath{\left[\mathrm{Na}/\mathrm{Fe}\right]}}

\newcommand{\bafe}{\ensuremath{\left[\mathrm{Ba}/\mathrm{Fe}\right]}}

\newcommand{\GIRAFFE}{{\tt GIRAFFE}}
\newcommand{\ATLAS}{{\tt ATLAS9}}

\newcommand{\MULTI}{{\tt MULTI}}

\begin{document} 

\title{Abundance of barium in the atmospheres of red giants \\in the Galactic globular cluster NGC~104 (47~Tuc)\thanks{Based on observations obtained at the European Southern Observatory (ESO) Very Large Telescope (VLT) at Paranal Observatory, Chile.}}

\author
	{
	V.~Dobrovolskas\inst{1}
	\and
	E.~Kolomiecas\inst{1}
	\and
	A.~Ku\v{c}inskas\inst{1}
	\and
	J.~Klevas\inst{1}
	\and
	S.~Korotin\inst{2}
	}

\institute
	{Institute of Theoretical Physics and Astronomy, Vilnius University, Saul\.{e}tekio al. 3, Vilnius, LT-10257, Lithuania\\
	\email{vidas.dobrovolskas@tfai.vu.lt}\\
	\and
	Crimean Astrophysical Observatory, Nauchny 298409, Crimea
	}

   \date{Received ; accepted}

% \abstract{}{}{}{}{} 
% 5 {} token are mandatory
 
  \abstract
  % context heading (optional)
  % {} leave it empty if necessary  
   {While most (if not all) Type~I Galactic globular clusters (GGCs) are characterised by spreads in the abundances of light chemical elements (e.g. Li, N, O, Na, Mg, Al), it is not yet well established whether similar spreads may exist in s-process elements as well.}
  % aims heading (mandatory)
   {We investigated the possible difference in Ba abundance between the primordial (1P) and polluted (2P) stars in the Galactic globular cluster (GGC) 47~Tuc (NGC~104). For this, we obtained homogeneous abundances of Fe, Na, and Ba in a sample of 261 red giant branch (RGB) stars which is the largest sample used for Na and Ba abundance analysis in any GGC so far.}
  % methods heading (mandatory)
   {Abundances of Na and Ba were determined using archival \GIRAFFE/{\tt VLT} spectra and 1D~non-local thermodynamic equilibrium (NLTE) abundance analysis methodology. 
   }
  % results heading (mandatory)
   {Contrary to the finding of \citet{GLS13}, we did not detect any significant Ba--Na correlation or 2P--1P Ba abundance difference in the sample of 261 RGB stars in 47~Tuc. This corroborates the result of \citet{dorazi10} who found no statistically significant Ba--Na correlation in 110 RGB stars in this GGC. The average barium-to-iron ratio obtained in the sample of 261 RGB stars, $\langle\bafe_{\rm 1D~NLTE}\rangle = -0.01\pm0.06$, agrees well with those determined in Galactic field stars at this metallicity and may therefore represent the abundance of primordial proto-cluster gas that has not been altered during the subsequent chemical evolution of the cluster.}
  % conclusions heading (optional), leave it empty if necessary
  {}
  %{The ratio of \lafe and \bafe abundances hint toward high-mass asymptotic giant branch stars as a possible polluters in 47~Tuc.}


\authorrunning{Dobrovolskas et al.}
\titlerunning{Abundance of barium in 47~Tuc}

   \maketitle
%
%-------------------------------------------------------------------

\section{Introduction}

Most if not all Galactic globular clusters contain multiple stellar populations which are characterised by different abundances of chemical elements, radial distributions in the given GGC, and kinematic properties \citep[see e.g.][for a review]{BL18}. It is usually assumed that a fraction of the GGC stars, so-called second-generation stars (2P), were enriched in certain chemical elements and depleted in others by some first-generation (1P) polluters. A number of enrichment scenarios have been proposed to explain the observed properties of the GGCs by involving various potential polluters, such as fast-rotating massive stars \citep[FRMS; e.g.,][]{DCS07}, asymptotic giant branch (AGB) stars \citep[e.g.,][]{DAVD16}, super-massive stars \citep[SMS, $\sim10^4\,{\rm M}_{\odot}$; e.g.,][]{DH14,GCK18}. 
Unfortunately, none of the proposed scenarios is capable of explaining chemical and kinematic differences between the 1P and 2P stars simultaneously \citep{BL18,GBC19}.

The 1P--2P differences in the abundances of chemical elements have been detected as correlations or anti-correlations between the abundances of light elements: Na--Li \citep{BPM07} correlation and O--Na, Mg--Al \citep{carretta09}, Li--O \citep{PBM05,SBP10} anti-correlations. Except for the so-called Type~II GGCs (see \citealt{MMK15,MMR19} for details), the majority of Galactic GGCs (Type~I) seem to be uniform in their Fe-group, as well as s- and r-process element abundances. 

As for the s-process elements, results of several studies have suggested that there may be some variation in their abundances in several Type~I GGCs, too. For example, \citet{GLS13} reported a possible Na--Ba correlation in the sample of 106 red horizontal branch (RHB) stars in NGC~104 (47~Tuc). While statistical significance of the possible correlation was found to be high, the authors have warned that, because of a small range in the [Ba/Fe] variation and [Na/O] correlation with the effective temperature along the horizontal branch (HB), their result needed to be confirmed in further studies for claiming a definite detection \citep{GLS13}. Some other studies have shown tentative signs for a possible correlation between the abundances of light and s-process elements in other Type~I GGCs, such as signatures for Sr--Na and Y--Na relations in M~4 (\citealt{SSG16,VG11}; but see also \citealt{DOCL13}). Also, our recent analysis of Zr abundance in 237 RGB stars in 47~Tuc suggests the existence of weak but statistically significant Na--Zr correlation and 2P--1P Zr abundance difference of $\Delta {\rm [Zr/Fe]_{\rm 2P-1P}}\approx0.06$ \citep{KDK21}.
The existence of such correlations would indicate that the light and s-process elements should have been produced in the same polluters that have enriched the 2P stars in some elements and depleted in others. Unfortunately, the data obtained so far is inconclusive thus further analysis of more s-process elements in the larger samples of GGC stars would be desirable.

With an aim of shedding more light on the possible 1P--2P differences in s-process element abundances, in this study we determined Ba abundances in 261 RGB stars in 47~Tuc, a Type~I GGC \citep{MMR19}. To the best of our knowledge, this is the only GGC in which a tentative detection of statistically significant relation between the light and s-process element abundances has been reported. Therefore, our primary goal was to verify whether the Na--Ba correlation and/or 1P--2P Ba abundance difference could be detected in our RGB star sample which is the largest one in which Ba abundance was determined in any GGC so far. 

The paper is structured as follows: after a short introduction we provide a brief description of the observational data (Sect.~\ref{sect:obs-data}) and abundance determination methodology (Sect.~\ref{sect:abnd-analysis}). The obtained results are discussed in Sect.~\ref{sect:results} and the main findings are summarized in the Conclusions.

\section{Observational data\label{sect:obs-data}}

In this work we studied a sample of RGB stars using the high-resolution spectra that were obtained with \GIRAFFE/{\tt VLT} during three observing programs, 072.D-0777(A), PI: Fran\c{c}ois; 073.D-0211(A), PI: Carretta; and 088.D-0026(A), PI: McDonald (Table~\ref{tab:obs-journal}). Individual spectra were retrieved for the analysis from the ESO Advanced Data Products Archive\footnote{http://archive.eso.org/wdb/wdb/adp/phase3\_spectral/form}. Our stellar sample overlaps with that utilized by \citet{KDK21} in their study of Zr in 47~Tuc (228 stars in common) but it also includes 33 additional targets that were not analysed in \citet{KDK21}.

Median-averaged sky spectrum was obtained from the dedicated sky fibres and subsequently was subtracted from the individual spectra of all target stars.
In the case of the program 088.D-0026(A) where three exposures for the same targets were available, spectra were co-added to increase signal-to-noise ratio, $S/N$, with the latter determined at the continuum level in the vicinity of the investigated \ion{Na}{i} and \ion{Ba}{ii} lines ($\sim 619.7$\,nm). The $S/N$ ratio of the final spectra varied from $S/N\sim220$ for the targets at $\teff \sim 4200$\,K to $S/N \sim 60$ at $\teff \sim 4700$\,K. All target spectra were continuum normalized using the \texttt{splot} task under {\tt IRAF} \citep{T86}. Radial velocities were determined using the \texttt{fxcor} task in {\tt IRAF} and cross-correlation technique. For this, we used synthetic spectrum computed with the model atmosphere having $\teff=4500$\,K, $\logg=1.90$, and $\feh=-0.7$ to represent a typical RGB star in 47~Tuc. 
The obtained radial velocities were then used to shift the wavelengths to the rest frame with the \texttt{dopcor} task in {\tt IRAF}.

Only RGB stars were selected for the abundance analysis, by discarding HB and AGB stars based on their position in the $V-(V-I)$ colour magnitude diagram. Before performing abundance analysis, we verified that all targets are members of 47~Tuc which was done based on their proper motions taken from the Gaia EDR3 catalogue \citep{G21}. Following \citet{MMM18}, we required that the proper motions of the cluster member stars did not deviate by more than 1.5\,mas\,yr$^{-1}$ from the mean cluster proper motion, $\mu_{\rm RA} = 5.25$\,mas\,yr$^{-1}$ and $\mu_{\rm DEC} = -2.53$\,mas\,yr$^{-1}$ \citep{BHS19}.

\begin{table*}[tb]
	\small
	\begin{center}
		\caption{Spectroscopic data used in this work.
			\label{tab:obs-journal}}
		\begin{tabular}{cccccccc}
			\hline\hline
			\noalign{\smallskip}
			Programme       & Date of      & Setting & $\lambda_{\mathrm{central}}$, & R  & Exposure, &  Number of \\
			& observations &         & nm                            &    & s         &    stars   \\
			\hline\noalign{\smallskip}
			072.D-0777(A)   & 2003-10-21   & HR13    & 627.3     & 22\,500      & 1500           &   112       \\
			& 2003-10-21   & HR13    & 627.3     & 22\,500      & 3600           &   121       \\
			073.D-0211(A)   & 2004-07-07   & HR13    & 627.3     & 22\,500      & 1600           &   113       \\
			088.D-0026(A)   & 2011-11-26   & HR13    & 627.3     & 22\,500      & $3\times700$   &   113       \\
			& 2011-11-26   & HR14A   & 651.5     & 17\,740      & $3\times340$   &   113       \\
			\hline
		\end{tabular}
	\end{center}
\end{table*}

Stars common to several programs were identified by cross-matching their coordinates within the 1\,arcsec radius. This resulted in 
14 stars in common to the 073.D-0211(A) and 088.D-0026(A) samples, 23 stars to 072.D-0777(A) and 088.D-0026(A), 17 stars to 072.D-0777(A) and 073.D-0211(A), and 5 stars in common to all three samples (Table~\ref{app-tab:common-stars}). Abundances of these stars were computed by averaging those obtained using spectra of individual observing programs.

The final target sample consisted of 261 unique RGB stars for which we determined Fe, Na, and Ba abundances. Our RGB sample is therefore somewhat larger than that of \citet[][237 objects]{KDK21} because it includes stars with $\teff>4800$\,K which were discarded from their analysis.

\begin{figure}[tb]
	\centering
	%	\resizebox{\hsize}{!}{\includegraphics{Fig-47Tuc-hrd.eps}}
	\includegraphics[width=8cm]{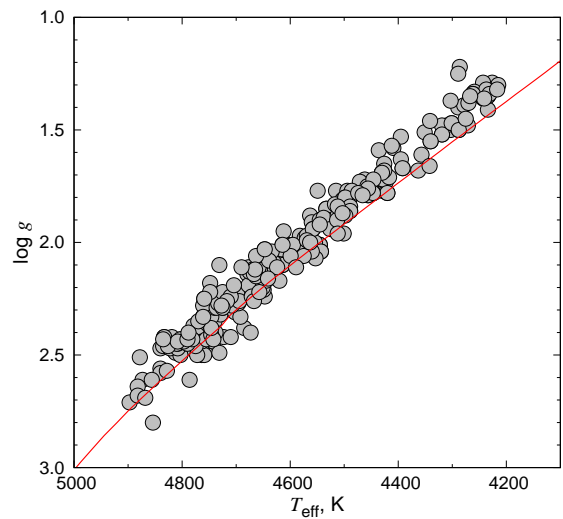}
	\caption{Target RGB stars in the $\teff-\log g$ plane. The red line is Yonsei-Yale \citep{YY2} isochrone (12 Gyr, $\mh = -0.68$, ${\rm [\alpha/Fe]}=+0.4$).}
	\label{fig:47Tuc-hrd}
\end{figure}

\begin{table*}
	\begin{center}
		\small
		\caption{Atomic parameters of \ion{Na}{i} and \ion{Ba}{ii} lines used in this study.
			\label{tab:line-list}}
		\begin{tabular}{llcrllc}
			\hline\hline
			\noalign{\smallskip}
			Element & $\lambda$, nm  & $\chi$, eV & log \textit{gf}$^{\mathrm{a}}$ & log $\gamma_{rad}$$^{\mathrm{b}}$ & log $\frac{\gamma_4}{N_e}$$^{\mathrm{c}}$ & log $\frac{\gamma_6}{N_H}$$^{\mathrm{d}}$ \\
			\hline\noalign{\smallskip}
			\ion{Na}{i}  & 615.4225 & 2.102 & $-1.547$ & 7.85                & $-4.39$     & $-7.28$               \\
			\ion{Na}{i}  & 616.0747 & 2.104 & $-1.246$ & 7.85                & $-4.39$     & $-7.28$               \\
			\ion{Ba}{ii} & 614.1730 & 0.704 & $-0.076$ & 8.20                 & $-5.48$     & $-7.47$   \\
			\ion{Ba}{ii} & 649.6910 & 0.604 & $-0.377$ & 8.10                 & $-5.48$     & $-7.47$   \\
			\hline
		\end{tabular}
	\end{center}
	\begin{list}{}{}
		\small
		\item[$^{\mathrm{a}}$] \citet{WM80}; $^{\mathrm{b}}$ \citet{MB96}; $^{\mathrm{c}}$ \citet{KRP00}; $^{\mathrm{d}}$ \citet{KMG11}
	\end{list}
\end{table*}

\subsection{Atmospheric parameters\label{subsect:atm-par}}

Atmospheric parameters for the majority of targets (228 stars) were determined by \citet{KDK21}. For the remaining 33 stars we used the same procedure as utilised by the latter authors. With this approach, effective temperatures were determined using photometry from \citet{BS09} and color-\teff\ calibration of \citet{RM05}. The obtained values were verified by checking trends in the iron abundance -- lower excitation potential plane. In most cases, the slopes were consistent with zero thereby suggesting a good agreement of photometric and spectroscopic \teff\ values.

Surface gravities were obtained by using a classical relation between stellar mass, luminosity, effective temperature, and surface gravity. This decision was made because only a few \ion{Fe}{ii} lines were available in the stellar spectra (typically, 1--4 lines). To compute gravities from photometry, we assumed an identical mass of 0.87\,M$_{\odot}$ for all RGB stars investigated in this work, as indicated by the Yonsei-Yale isochrone of 12\,Gyr and $\mh=-0.68$ \citep{YY2}. Bolometric luminosities were computed using relation between the bolometric correction, \teff, and metallicity from \citet{alonso99}. 

Microturbulence velocities were obtained by enforcing zero trend in the iron abundance -- equivalent width plane while keeping photometric effective temperature fixed. In this procedure strong lines were neglected ($W>15$\,pm) because of their reduced sensitivity to microturbulence velocity. Due to small number (17-28) of iron lines, the estimated accuracy of microturbulence velocity was $\approx0.2$\,km\,s$^{-1}$ and represents the main source of uncertainty in the determined Ba abundances (Section~\ref{subsect:abund-err}).

The location of the target RGB stars in the $\teff-\log g$ plane is shown in Fig.~\ref{fig:47Tuc-hrd}. Atmospheric parameters of the individual target stars are listed in Table~\ref{app-tab:abund-indiv-stars}.

\section{Abundance analysis\label{sect:abnd-analysis}}

Because the stellar sample used in the present study overlaps with that investigated by \citet{KDK21}, abundances of Fe and Na in targets common to the two studies (228 stars) were taken from \citet{KDK21}. In addition, we determined Fe and Na abundances in 33 RGB stars and Ba abundances in 261 stars. In both studies abundance analysis of Fe and Na was performed following strictly identical procedures. Abundances of Na and Ba were determined using \ATLAS\ model atmospheres \citep{Kur93,SBC04,S05} and spectral synthesis, under the assumption of non-local thermodynamic equilibrium (NLTE), while local thermodynamic equilibrium (LTE) was assumed in the analysis of Fe. Procedures involved in the abundance determination are described below.

\subsection{Reference abundances in the Sun\label{subsect:ref-abnd}}

Reference abundances in the Sun were determined using the Kitt Peak Solar Flux atlas \citep{KFB84}. To briefly summarize, Fe abundance was obtained using the equivalent width method, with the latter determined by fitting the Voigt profiles to the observed spectral line profiles. The \ion{Fe}{i} and \ion{Fe}{ii} line list used for abundance determination is provided in Table~\ref{app-tab:iron-list}. Oscillator strength data were obtained from the VALD3 database \citep{RPK15}.

Microturbulence velocity was obtained by iteration until there was zero trend of iron abundance with the line strength. Our determined value of \vmicro = 0.93 km\,s$^{-1}$ is in good agreement with the typical value of 0.9-1.0 km\,s$^{-1}$ available in the literature \citep{Doyle14}. The average solar iron abundance obtained with the determined microturbulence velocity was \textit{A}(Fe) = $7.55 \pm 0.01$ ($\sigma = 0.06$;  here $\pm0.01$ is the error of the mean, $\sigma$ denotes standard deviation due to the line-to-line abundance variation) which was obtained from 29 lines with \textit{W}<10.5\,pm (Table~\ref{app-tab:iron-list}). Although the scatter is significant, the average abundance is in good agreement with those available in the literature, e.g. $A({\rm Fe})=7.51\pm0.06$ from \citet{CLS11}.

For the solar Na abundances, we used $A({\rm Na})= 6.17$ which was determined with the NLTE spectral line synthesis methodology (Sect.~\ref{subsect:Na-determ}) and by using the same \ion{Na}{i} lines as utilised in the analysis of the target stars in 47~Tuc. This value agrees well with those obtained in the earlier studies, e.g. $A({\rm Na})_{\odot}=6.21\pm0.04$ derived by \citet{SAG15_Na}.

Solar abundance of Ba was obtained by utilising the NLTE methodology (Sec.~\ref{subsect:Ba-determ}). The determined value, \textit{A}(Ba) = 2.17, agrees reasonably well with $A({\rm Ba})_{\rm 1D~NLTE}=2.07$ and $A({\rm Ba})_{\rm 3D~NLTE}=2.28$ derived by \citet{GBC20} and coincides with the meteoritic value of $2.17\pm0.02$ from \citet{L21}.

\begin{figure*}[tb]
	\centering
	%%	\resizebox{\hsize}{!}{\includegraphics{Fig-47Tuc-na-line-fit.eps}}
	{\includegraphics[width=16cm]{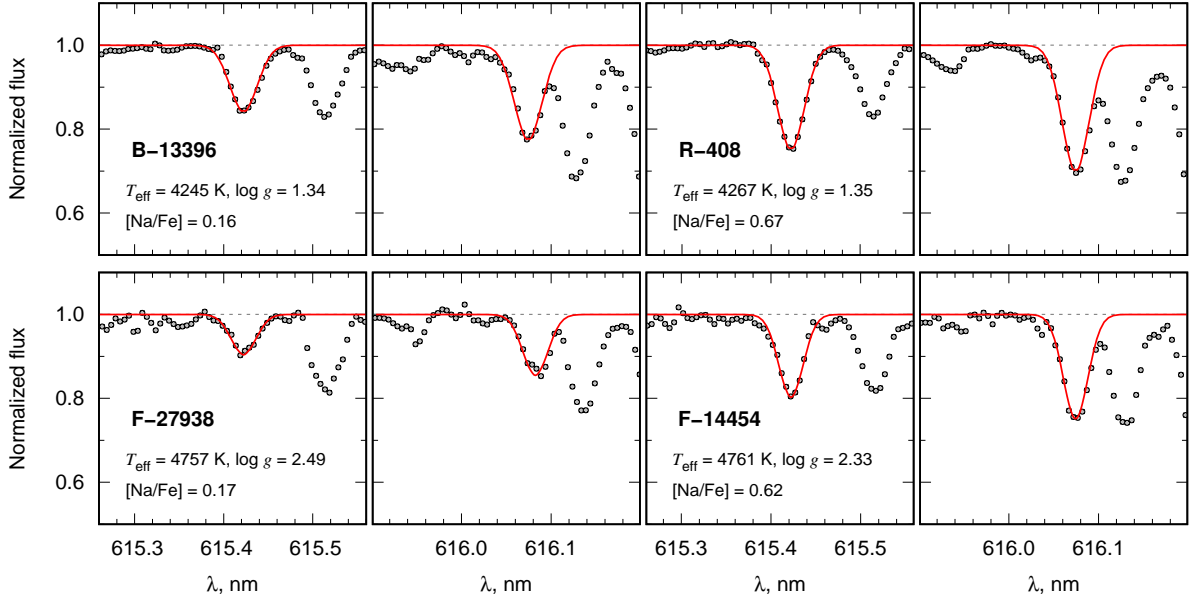}}
	\caption{Examples of the observed (black dots) and best-fitted synthetic \ion{Na}{i} line profiles (red lines) in the GIRAFFE spectra of the target stars characterised by different Na abundances and effective temperatures ($\teff\approx4250$\,K, top row; $\teff\approx4750$\,K, bottom row). The continuum level is shown as the gray dashed line.
	}
	\label{fig:47Tuc-na-line-fit}
\end{figure*}

\begin{figure}[tb]
	\centering
	%%	\resizebox{\hsize}{!}{\includegraphics{Fig-47Tuc-ba-line-fit.eps}}
	{\includegraphics[width=8cm]{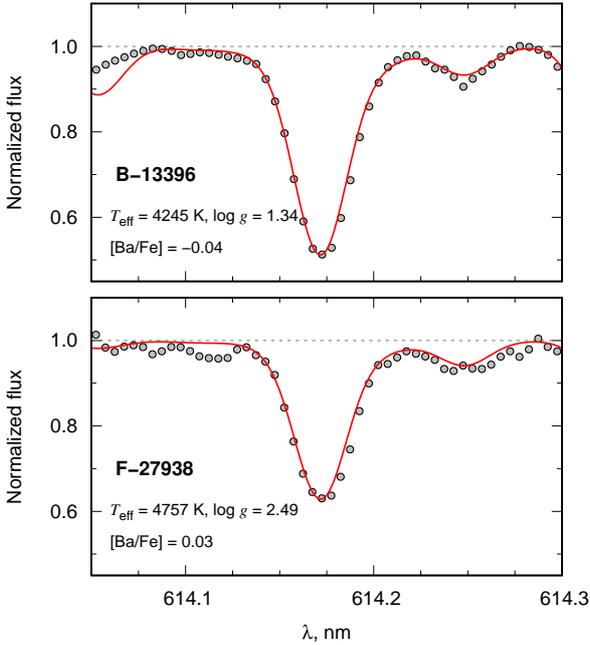}}
	\caption{Examples of the observed (black dots) and best-fitted synthetic spectra in the vicinity of \ion{Ba}{ii} lines (red lines) in the GIRAFFE spectra of the target stars characterised by different effective temperatures ($\teff\approx4250$\,K, top panel; $\teff\approx4750$\,K, bottom panel). The continuum level is shown as the gray dashed line.}
	\label{fig:47Tuc-ba-line-fit}
\end{figure}

\subsection{Determination of Fe abundance\label{subsect:Fe-determ}}

Abundances of Fe in the target RGB stars were obtained using a set of 17--28 \ion{Fe}{i} lines ($612.79-691.67$\,nm), with the line lower level excitation potentials in the range of 2.18--4.61\,eV (Table~\ref{app-tab:iron-list}). 
The mean value obtained in the sample of 261 stars was $\langle{\rm [Fe/H]}\rangle=-0.75\pm 0.05$ (the error denotes standard deviation due to the star-to-star abundance scatter) which agrees well with the values determined in the earlier studies (cf. $\langle{\rm [Fe/H]}\rangle=-0.74\pm0.05$, \citealt[][114 RGB stars]{carretta09}). The  Fe abundances obtained in 261 target RGB stars are listed in Table~\ref{app-tab:abund-indiv-stars}.

\subsection{Determination of Na abundance\label{subsect:Na-determ}}

Two \ion{Na}{i} lines located at 615.42\,nm and 616.07\,nm were used in the determination of Na abundances (Table~\ref{tab:line-list}). The model atom of Na used in this work was the same as described in \citet{dobrovolskas14}. It consisted of the first 20 energy levels of \ion{Na}{i} and the ground level of \ion{Na}{ii}, with a total of 46 radiative transitions taken into account. The rate coefficients of collisions with hydrogen atoms for the lower 9 levels of \ion{Na}{i} were taken from \citet{BBD10}. 
Examples of the best-fitted \ion{Na}{i} line profiles in the Na-poor and Na-rich stars are shown in Fig.~\ref{fig:47Tuc-na-line-fit}, while the Na abundances determined in all target stars are plotted in the middle row of Fig.~\ref{fig:abn-vs-atmpar} and are listed in Table~\ref{app-tab:abund-indiv-stars}. 
The Na NLTE--LTE abundance corrections, $\Delta_{\mathrm{\mathrm{1D\,NLTE-LTE}}}=A(\mathrm{Na})_{\mathrm{1D\,NLTE}}-A(\mathrm{Na})_{\mathrm{1D\,LTE}}${}, were in the range $-0.07...-0.27$, depending on the strength of the spectral line and effective temperature of the target star (Table~\ref{tab:nlte-abund-corr}).

\begin{table}
	\begin{center}
		\caption{1D NLTE-LTE abundance corrections for Na and Ba.
			\label{tab:nlte-abund-corr}}
		%\resizebox{\hsize}{!}{
		\begin{tabular}{cccc}
			\hline\hline
			\noalign{\smallskip}
			Element      & $\lambda$,\,nm &    \textit{W},\,pm     & $\Delta_{\mathrm{1D\, NLTE-LTE} }$ \\
			\hline\noalign{\smallskip}
			\multicolumn{4}{c}{$\teff=4250$\,K, $\logg = 1.35$} \\
			\multicolumn{4}{l}{\textit{Na-poor}} \\
			\ion{Na}{i}  & 615.42         &      5.7      & $-0.08$             \\
			\ion{Na}{i}  & 616.07         &      8.7      & $-0.14$             \\
			\ion{Ba}{ii} & 614.17         &     19.0      & $-0.06$             \\
			\multicolumn{4}{l}{\textit{Na-rich}} \\
			\ion{Na}{i}  & 615.42         &      9.6      & $-0.19$             \\
			\ion{Na}{i}  & 616.07         &     11.8      & $-0.27$             \\
			\multicolumn{4}{c}{$\teff=4750$\,K, $\logg = 2.40$} \\
			\multicolumn{4}{l}{\textit{Na-poor}} \\
			\ion{Na}{i}  & 615.42         &      3.6      & $-0.07$             \\
			\ion{Na}{i}  & 616.07         &      5.4      & $-0.08$             \\
			\ion{Ba}{ii} & 614.17         &     14.8      & $-0.10$             \\
			\multicolumn{4}{l}{\textit{Na-rich}} \\
			\ion{Na}{i}  & 615.42         &      6.3      & $-0.12$             \\
			\ion{Na}{i}  & 616.07         &      8.7      & $-0.15$             \\
			\hline
		\end{tabular}
		%}
	\end{center}
\end{table}

\subsection{Determination of Ba abundance\label{subsect:Ba-determ}}

As in the case of Na, abundances of Ba were determined using spectral synthesis approach, under the assumption of NLTE. Two \ion{Ba}{ii} lines located at 614.17\,nm and 649.69\,nm were used for the analysis whenever possible though in most cases only the 614.17\,nm line was available for the analysis (Table~\ref{tab:line-list}). The level departure coefficients and spectral line profiles were computed using the version of \MULTI\ code \citep{carlsson} modified by \citep{korotin99}. To take into account the fact that 614.17\,nm line is blended with \ion{Fe}{i} line, the spectra were computed with the {\tt SynthV} code \citep{tsymbal96}. For this, we used NLTE departure coefficients of Ba computed with the \MULTI\ code while the \ion{Fe}{i} line was synthesized under the assumption of LTE. The model atom of Ba used in this work was the same as described in \citet{andrievsky09}. It consisted of 31 levels of \ion{Ba}{i}, 101 levels of \ion{Ba}{ii} ($n<50$), and the ground level of \ion{Ba}{iii}. In total, 91 bound-bound transitions between the lowest 28 energy levels ($n<12$, $l<5$) of \ion{Ba}{ii} were taken into account. Fine structure was included for the levels 5d$^2$D and 6p2P$^0$, according to the prescription given in \citet{andrievsky09}. Following the latter study, the hyperfine structure of the 649.69\,nm line was approximated using three components (lines).

\begin{figure}[tb]
	\resizebox{\hsize}{!}{\includegraphics{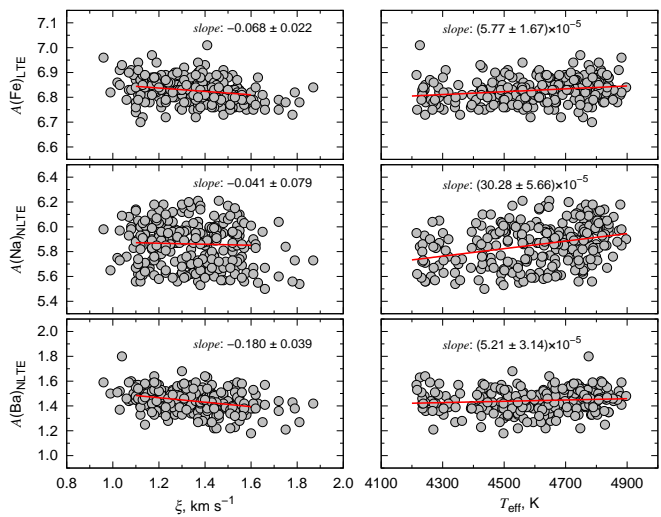}}
	\caption{Fe (top row), Na (middle row), and Ba (bottom row) abundances plotted against the microturbulence velocity (left column) and effective temperature (right column). In all panels the linear fits to the data are shown as the solid red lines.}
	\label{fig:abn-vs-atmpar}
\end{figure}

\begin{table}
	\begin{center}
		\caption{Typical Fe, Na, and Ba abundance measurement errors.
		}
		\label{tab:abund-errors}
		\resizebox{\hsize}{!}{%
			\begin{tabular}{cccccccc}
				\hline\hline
				\noalign{\smallskip}
				Element & $\sigma$(\textit{T}$_{\mathrm{eff}}$)  & $\sigma$(log\textit{g})  & $\sigma$($\xi_\mathrm{t}$)  & $\sigma$(cont)  & $\sigma$(fit)  &$\sigma$$_{A{\rm (X)}}$(total)  \\
				&    dex     & dex                            &  dex  & dex & dex  & dex            \\
				\hline\noalign{\smallskip}
				\multicolumn{7}{c}{$\teff=4250$\,K, $\logg = 1.35$} \\
				Fe	&	0.03	&	0.02	&	0.12	&	0.10	&	0.02	&	0.16	\\
				Na &   0.06	   &	0.01	&	0.05	&	0.03	&	0.04	&	0.09	\\
				Ba	&	0.01	&	0.04	&	0.10	&	0.04	&	0.03	&	0.12	\\
				\multicolumn{7}{c}{$\teff=4750$\,K, $\logg = 2.40$} \\
				Fe	&	0.04	&	0.02	&	0.11	&	0.10	&	0.05	&	0.16	\\
				Na &   0.06	   &	0.01	&	0.03	&	0.03	&	0.05	&	0.09	\\
				Ba	&	0.02	&	0.02	&	0.08	&	0.05	&	0.06	&	0.12	\\
				\hline
			\end{tabular}
		}
	\end{center}
\end{table}

Only the 088.D-0026(A) sample contained spectra where both \ion{Ba}{ii} lines were available for every target star. In this case, the final Ba abundance value was taken as an average of estimates obtained from the two lines. In all other cases Ba abundance was determined using a single 614.17\,nm line. Since \ion{Ba}{ii} lines are strong, we did not take into account their possible contamination by CN lines and their influence on the Ba abundance determination. We nevertheless estimate that this effect should not exceed 0.01\,dex. The obtained Ba abundances are shown in the bottom row of Fig.~\ref{fig:abn-vs-atmpar} and are listed in Table~\ref{app-tab:abund-indiv-stars}. Typical examples of the 1D NLTE synthetic spectrum fits to the \ion{Ba}{ii} 614.17\,nm spectral line in the target stars characterised by different effective temperatures are provided in Fig.~\ref{fig:47Tuc-ba-line-fit}. The $\Delta_{\mathrm{1D\,NLTE-LTE}}$ abundance corrections for Ba were in the range $-0.06...-0.10$, increasing in magnitude at higher effective temperatures (Table~\ref{tab:nlte-abund-corr}).

The average barium-to-iron abundance ratio determined in the sample of 261 was $\langle\bafe_{\rm 1D~NLTE}\rangle = -0.01\pm0.06$ where the error is standard deviation due to star-to-star abundance variation.

\subsection{Errors in the determined Fe, Na, and Ba abundances\label{subsect:abund-err}}

Uncertainties in the determined Fe, Na, and Ba abundances were determined using the prescription given in \citet{cerniauskas18}, by utilising \ATLAS\ models computed with $\teff=4250$\,K, $\log g=1.35$; and $\teff=4750$\,K, $\log g=2.40$. In essence, we first estimated errors in the effective temperature, surface gravity, microturbulence velocity, continuum determination and line profile fitting. For all sample stars, we assumed identical uncertainties in the effective temperature, surface gravity, and microturbulence velocity, which were 80 K, 0.1 dex, and 0.1 km\,s$^{-1}${}, respectively. Error of the continuum placement was determined by computing standard deviation of the continuum, $\sigma_{\rm cont}$, in the vicinity of the investigated spectral lines and then changing continuum level by $\pm 1\sigma_{\rm cont}$ from the adopted level and re-determining abundances. The fit error was estimated by computing standard deviation of the difference between the observed and best-fitted spectral line profiles. Again, the synthetic spectrum was changed by $\pm 1\sigma${} and the resulting difference in the abundance was computed. These individual errors were then added in quadratures and were used further as uncertainties in the determined abundances of Fe, Na, and Ba. The typical abundance errors are provided in Table~\ref{tab:abund-errors}.

\section {Results and discussion\label{sect:results}}

\subsection{Possible Ba--Na anti-correlation in the RGB stars in 47~Tuc?}

\begin{figure}[tb]
	\centering
	\includegraphics[width=7cm]{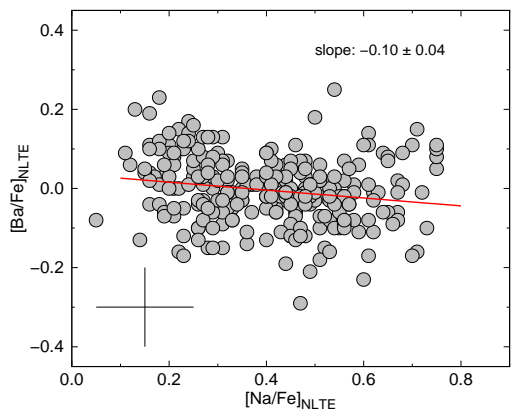}
	%	\resizebox{\hsize}{!}{\includegraphics{Fig-Ba-vs-Na.eps}}
	\caption{[Ba/Fe] abundance ratios determined in 261 RGB stars in 47~Tuc, plotted versus their [Na/Fe] abundance rations. Typical abundance error bars are shown in the bottom left corner. The linear fit to the data is shown as the solid red line.}
	\label{fig:ba-abnd}
\end{figure}

The barium-to-iron abundance ratios, \bafe, determined in 261 RGB stars show a weak but statistically significant anti-correlation with the sodium-to-iron abundance ratios, \nafe{} (Fig~\ref{fig:ba-abnd}), with the Pearson's correlation coefficient $r_{\mathrm P} = -0.18$. Assuming a null-hypothesis that there is no correlation between the two abundance ratios, the Student's $t$-test gives a probability of $p_{\mathrm P} = 2.9 \times 10^{-3}$ that such $r_{\mathrm P}$ value could be obtained by chance (Table~\ref{tab:corr-coef}). The Spearman's rank correlation coefficient and the $p$-value are, respectively, $\rho_{S} = -0.20$ and $p_{\mathrm S} = 1.3 \times 10^{-3}$, while for the Kendall's rank correlation the two values are $\tau_{K} = -0.14$ and $p_{\mathrm K} = 6.9 \times 10^{-4}$, respectively.

\begin{figure}[tb]
	\centering
	\includegraphics[width=7cm]{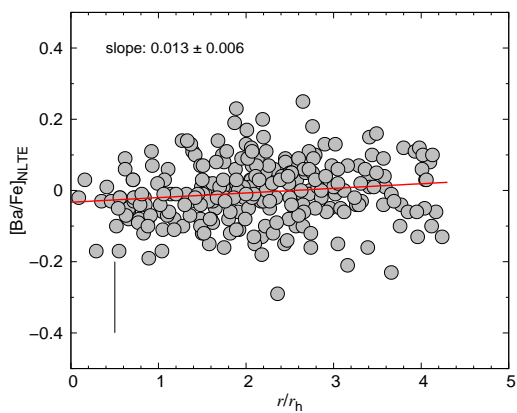}
	%	\resizebox{\hsize}{!}{\includegraphics{Fig-bafe-vs-rad-dist.eps}}
	\caption{[Ba/Fe] abundance ratios obtained in 261 RGB stars in 47~Tuc, plotted versus relative radial distance from the cluster center. Cluster half-mass radius, $r_{\mathrm h} = 174$\,arcsec, was taken from \citet{trager93}. Linear fit to the data is marked as the solid red line. Typical abundance error is shown as the vertical bar in the bottom left corner.}
	\label{fig:ba-abnd-raddist}
\end{figure}

\begin{table}
\begin{center}
\caption{The $\bafe-\nafe$ abundance correlation coefficients and their in the full sample of 261 RGB stars in 47~Tuc and sub-samples divided into $\Delta\teff=100${} K-wide bins. $N$ is the number of stars in the given (sub-)sample.
\label{tab:corr-coef}}
\resizebox{\hsize}{!}{
\begin{tabular}{cccccc}
\hline\hline
\noalign{\smallskip}
\multicolumn{2}{c}{Pearson}         &  \multicolumn{2}{c}{Spearman}     &   \multicolumn{2}{c}{Kendall}     \\
    $r$    &    $p$                 &   $\rho$  &    $p$                &  $\tau$   &      $p$              \\
\noalign{\smallskip}
\hline
\noalign{\smallskip}
\multicolumn{6}{c}{Full sample, N = 261} \\
  $-0.18$  &  $2.93\times 10^{-3}$  &  $-0.20$  &  $1.30\times 10^{-3}$  &  $-0.14$ &  $6.88\times 10^{-4}$ \\
\noalign{\smallskip}
\hline
\noalign{\smallskip}
\multicolumn{6}{c}{$4200 \leq \teff < 4300$, N = 23} \\
  $-0.11$  & 0.61                   & $-0.09$   &  0.68                  &  $-0.06$ &  0.68                 \\
\noalign{\smallskip}
\multicolumn{6}{c}{$4300 \leq \teff < 4400$, N = 15} \\
  $-0.15$  & 0.60                   & $-0.17$   &  0.54                  &  $-0.16$ &  0.42                 \\
\noalign{\smallskip}
\multicolumn{6}{c}{$4400 \leq \teff < 4500$, N = 35} \\
  $-0.38$  & 0.024                  & $-0.34$   &  0.048                 &  $-0.24$ &  0.043                \\
\noalign{\smallskip}
\multicolumn{6}{c}{$4500 \leq \teff < 4600$, N = 42} \\
  $-0.15$  & 0.35                   & $-0.15$   &  0.32                  &  $-0.11$ &  0.31                 \\
\noalign{\smallskip}
\multicolumn{6}{c}{$4600 \leq \teff < 4700$, N = 48} \\
  $-0.41$  & $3.77\times 10^{-3}$   & $-0.41$   &  $3.19\times 10^{-3}$  & $-0.28$  &  $4.87\times 10^{-3}$ \\
\noalign{\smallskip}
\multicolumn{6}{c}{$4700 \leq \teff < 4800$, N = 66} \\
 $-0.11$   & 0.39                   & $-0.16$   &  0.20                  &  $-0.12$ &  0.15                 \\
\noalign{\smallskip}
\multicolumn{6}{c}{$4800 \leq \teff < 4800$, N = 32} \\
 0.15      & 0.42                   & 0.04      &  0.83                  &  0.03    &  0.84                 \\
\hline\hline
\end{tabular}
}
\end{center}
\end{table}

We also checked whether such anti-correlation may exist between the determined [Ba/Fe] abundance ratios and the distance of the target RGB stars from the cluster center. For this, we adopted the coordinates of the 47~Tuc center, $\alpha_0(J2000)$ = 6.023625\,deg and $\delta_0(J2000)$ = -72.081276\,deg, from \citet{BHS19} and the half-mass radius of $r_{\mathrm h} = 174$\,arcsec from \citet{trager93}. Our data suggest a weak correlation in the ${\rm [Ba/Fe]}-r_{\mathrm h}$ plane, with $r_{\rm P} = 0.14$ and $p_{\rm P} = 0.019$ (Fig.~\ref{fig:ba-abnd-raddist}); similar values were obtained for the Spearman's and Kendall's rank correlation coefficients.

On the one hand, these results may suggest that in 47~Tuc the 1P stars (lower Na abundances) are more abundant in Ba than the 2P stars (higher Na abundances; Fig.~\ref{fig:ba-abnd}). Because the 2P stars tend to concentrate towards the cluster center  in this and other GGCs \citep[e.g.][]{KDB14}, this should lead to a weak correlation in the ${\rm [Ba/Fe]}-r_{\mathrm h}$ plane. This is exactly what our data suggest (Fig.~\ref{fig:ba-abnd-raddist}). However, this result would be difficult to explain from the nucleosynthesis point of view. It is well known that Ba could be synthesised either in the low- and intermediate-mass AGB stars via the main s-process or during the central He-burning phase in the massive rotating stars. However, the trends suggested by our data would require a mechanism which would destroy Ba in the 2P stars instead of synthesising it like, e.g. Na.

\begin{figure}[tb]
	\centering
	\includegraphics[width=7cm]{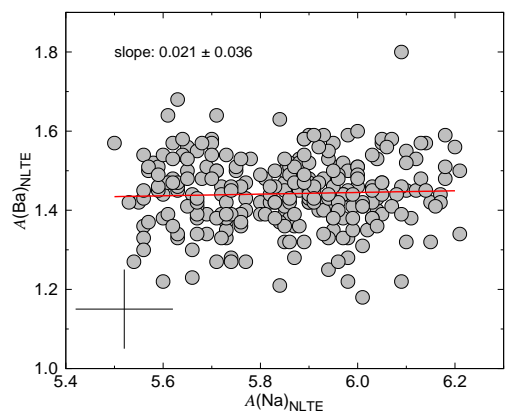}
%	\resizebox{\hsize}{!}{\includegraphics{Fig-Ba-vs-Na-abs.eps}}
	\caption{Absolute abundances of Ba and Na in the $A({\rm Ba})-A({\rm Na})$ plane. Typical abundance error bars are shown in the bottom left corner. The linear fit to the data is shown as the solid red line.}
	\label{fig:ba-abnd-abs}
\end{figure}

On the other hand, no Ba--Na anti-correlation is seen in the absolute abundance plane, $A({\rm Ba})-A({\rm Na})$ (Fig.~\ref{fig:ba-abnd-abs}). In this case we obtained $r_{\mathrm P} = 0.04$ and $p_{\mathrm P} = 0.52$, with similar results also for the Spearman's and Kendall's rank correlation coefficients: $\rho_{S} = 0.03$ and $p_{\mathrm S} = 0.63$ for the former, and $\tau_{K} = 0.03$ and $p_{\mathrm K} = 0.53$ for the latter. No trend was detected in the ${\rm [Ba/Fe]}-r_{\mathrm h}$ plane either, with $r_{\rm P} = -0.002\pm0.006$ and $p_{\rm P} = 0.984$. This therefore suggests that the anti-correlation in the ${\rm [Ba/Fe]}-{\rm [Na/Fe]}$ plane and correlation in the ${\rm [Ba/Fe]}-r_{\mathrm h} $ plane may be caused by the variation in Fe abundance alone.

A closer inspection of the data has shown that the Ba--Na anti-correlation in the $\bafe-\nafe$ plane has been detectable only in a single effective temperature interval (bin), $4600 \leq \teff < 4700$ \,K, with a possible weaker trend seen also in the $4400 \leq \teff < 4500$\,K bin (Table~\ref{tab:corr-coef}). No statistically significant correlation was detected in other $\Delta=100$\,K wide effective temperature bins. It therefore seems that the anti-correlation in the $4600 \leq \teff < 4700$ \,K bin was caused by several outlying stars with the somewhat higher and lower Ba abundances at the lowest and highest Na abundance ends, respectively (see Sect.~\ref{app-sect:ba-abund-teff46-47}). No residual correlation is seen if these outliers are excluded. 

It is also important to stress that abundances of all there elements analysed in this study show trends with the microturbulence velocity (Fig~\ref{fig:abn-vs-atmpar}). The most likely explanation for this is that microturbulence velocities were slightly underestimated in the hotter ($\teff\gtrapprox4700 K$) stars which are fainter and therefore their spectra are of lower quality. For these stars, a slight overestimate of the continuum during the determination of microturbulence velocities would lead to a systematically larger abundances determined from the weaker \ion{Fe}{i} lines with respect to those obtained from the stronger lines. This, in turn, would require a lower microturbulence velocity value to eliminate the trend of iron abundance with the equivalent width. Because the \ion{Ba}{ii} lines are the strongest of all three elements studied, it is not surprising that there is the strongest trend of abundance versus microturbulence velocity for barium. Thus, overestimated Ba abundances in stars with the lowest microturbulence velocities (i.e. those with the highest effective temperatures) could also contribute to the spurious $\bafe-\nafe$ anti-correlation.

Therefore, the most likely reason for the anti-correlation seen in the $\bafe-\nafe$ plane and correlation in the $\bafe-r_{\mathrm h}$ plane may be a random fluctuation in Ba abundances in several stars with the extreme Na abundances in the $4600 \leq \teff < 4700$ \,K effective temperature bin. This deviation in a single \teff\ bin can not be explained by e.g. the lower quality of the spectra, because for the hotter and therefore fainter stars there is no such correlation. Similarly, there is no plausible astrophysical reason why such correlation should exist in a single \teff\ bin only. We also note that no Ba--Na correlation was detected by \citet{dorazi10} in their analysis of 110 RGB stars in 47~Tuc which indirectly supports our assertion that the $\bafe-\nafe$ anti-correlation detected in our study is in fact spurious. This said, it would be nevertheless interesting to see whether the Ba--Na - or correlation between any other s-process and light elements - could be seen in other GGCs, especially given the tentative detection of Zr--Na correlation in the RGB stars of 47~Tuc suggested in the analysis of \citet{KDK21}.

\subsection{Average Ba abundance in the RGB stars in 47~Tuc}

Until now, Ba abundance in 47~Tuc has been determined in a number of studies. In one of the earlier attempts, \citet{JFB04} investigated Ba in the atmospheres of turn-off (TO) and subgiant (SGB) stars in 47~Tuc. The average Ba LTE abundance determined in eight SGB stars was \bafe=+0.35, with only slightly lower value of \bafe=+0.22 obtained in three TO stars. The average value derived for these stars is therefore slightly larger than the average abundance obtained in our study, given that we determined $\langle\bafe_{\rm 1D~NLTE}\rangle = -0.01\pm0.06$ and that the mean 1D~NLTE--LTE Ba abundance correction is $\sim -0.10$\,dex.

One of the most comprehensive Ba abundance studies in the GGCs was performed by \citet{dorazi10} who studied 1200 stars in 15 GGCs. In case of 47~Tuc they obtained a mean value of $\langle\bafe_{\rm 1D~LTE}\rangle = 0.15\pm0.06$ in the sample of 110 RGB stars (the error is standard deviation due to star-to-star abundance variation). Again, this value agrees well with that obtained in the present study if the NLTE--LTE correction ($\sim -0.10$\,dex) is taken into account. Similarly, a reasonable agreement is obtained with the average Ba LTE abundances obtained by \citet[][$\bafe=0.34\pm0.33$, 3 RGB stars]{worley10} and \citet[][$\bafe=0.32\pm0.05$, 13 RGB stars]{worley12} if the NLTE--LTE abundance correction is applied.

In their analysis of 13 RGB stars in 47~Tuc \citet{thygesen14} determined an average Ba NLTE abundance $\bafe=0.28\pm0.07$ which is significantly higher than that obtained in our study. A possible reason for this discrepancy is differences in the determined surface gravities and microturbulence velocities. Judging from the analysis of 6 stars that were available both in their and our samples we find that on average our microturbulence velocities and gravities were higher by 0.33\,km/s and 0.3\,dex, respectively. Correction for these differences would make our abundances higher by 0.20\,dex and would bring them into a  good agreement with the value obtained by \citet{thygesen14}.

\begin{figure}[tb]
	\resizebox{\hsize}{!}{\includegraphics{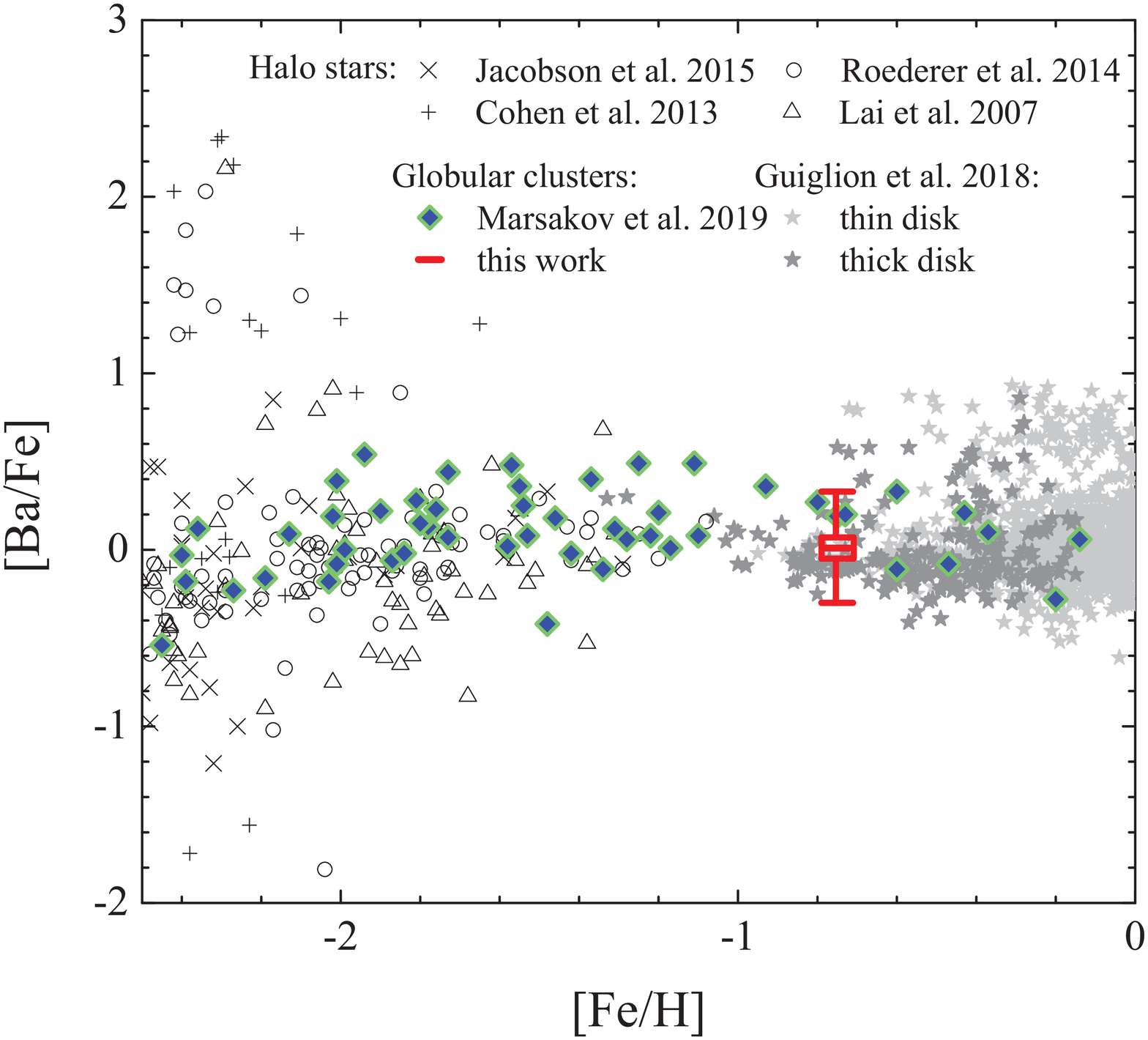}}
	\caption{Abundance of barium in the Galactic globular clusters (compilation by \citet{marsakov19}) and Galactic field stars (thin/thick disk stars: \citet{ambre18}, halo stars: \citet{LJB07,CCT13,RPT14,JKF15}). Average [Ba/Fe] ratio obtained in 261 RGB stars in 47~Tuc is marked by the red symbol where the 25th, 75th percentiles and the average values are indicated by the horizontal bars while the extent of the whiskers correspond to the minimum and the maximum abundance values observed in this cluster.
	}
	\label{fig:gc-ba-abnd}
\end{figure}

\subsection{Implications for the chemical evolution scenarios of 47~Tuc}

Results of Ba abundance analysis in 261 RGB stars suggest that there have been no enrichment or depletion of Ba in the 2P stars of 47~Tuc. Similarly, no Ba--Na correlation have been detected by \citet{dorazi10} in their analysis of 110 RGB stars. These findings contradict the result obtained by \citet{GLS13} who observed a weak but statistically significant Ba--Na correlation in the sample of 114 RHB stars in 47~Tuc. However, as was indicated by \citet{GSA15}, their detection of a possible Ba--Na correlation should be taken with caution because the [Na/O] correlation with effective temperature along the HB could partly or even fully explain the observed weak Ba--Na trend.

The obtained average barium-to-iron abundance ratio in 261 RGB stars in 47~Tuc, $\langle\bafe_{\rm 1D~NLTE}\rangle = -0.01\pm0.06$, agrees well with the abundances obtained at this metallicity in other GGCs, as well as with those determined in the Galactic disc and halo stars (Fig.~\ref{fig:gc-ba-abnd}). This, as well as the relatively narrow spread in Ba abundances of $\pm 0.06$\,dex (which can be fully accounted for by the Ba abundance determination error, $\sigma_{{\rm A(Ba)}}\approx0.12$\,dex), suggest that Ba abundance in 47~Tuc may reflect that of the primordial proto-cluster gas that has not been altered significantly (if at all) in the subsequent chemical enrichment during the evolution of the GGC.

In a recent analysis of Zr abundance in 237 RGB stars in 47~Tuc, \citet{KDK21} detected a weak but statistically significant Zr--Na correlation and 2P--1P Zr abundance difference of 0.06\,dex. Assuming that the 2P were indeed enriched in Zr but not in Ba would suggest that only some s-process elements have been produced by the polluters that have enriched the 2P stars in Na and modified abundances of other light elements such as Li, N, O, Mg, Al, as determined in the earlier studies \citep[see e.g.][]{BL18}. Theoretical yields from low- and intermediate-mass AGB stars predict that Zr and Ba can be produced in substantial amounts but they should be synthesised simultaneously \citep{CSP15}. Similarly, substantial amounts of Zr and Ba could be produced by explosive nucleosynthesis in massive rotating stars but, again, both elements should be produced simultaneously \citep{LC18}. Besides, in the latter scenario one would also expect some enrichment in the Fe-group and r-process elements which has not been observed in the Type~I GGCs until now \citep[e.g.][]{BL18,GBC19}. It would be therefore very desirable to verify whether correlations between the other s-process and light chemical elements may exist in 47~Tuc as well as in other Type~I GGCs.

\section{Conclusions\label{sect:conclusions}}

We present homogeneous abundances of Fe, Na, and Ba obtained in the sample of 261 RGB stars that belong to Galactic globular cluster (GGC) 47~Tuc. Contrary to the earlier finding of \citet{GLS13}, our results show no statistically significant variation of Ba abundance with that of Na thus suggesting that the primordial (Na-poor, 1P) and polluted (Na-rich, 2P) populations in this GGC are characterized by the same average Ba abundance, $\bafe_{\rm 1D~NLTE}=0.02\pm0.06$ (here the error is standard deviation due to star-to-star abundance variation). This is also supported by the analysis of \citet{dorazi10} who detected no Ba--Na correlation in a sample of 110 RGB stars in this GGC. Taken together with the recent finding of \citet{KDK21} who reported a detection of weak but statistically significant Zr--Na correlation in 237 RG stars in 47~Tuc, these results would indicate that in this GGC the 2P stars may have been enriched in certain s-process elements (i.e. Zr) but not in others (i.e. Ba). It would be therefore very desirable to verify whether such correlations between the abundances of s-process and light chemical elements may exist in other GGCs via the analysis of more s-process elements in large samples of GGC stars, preferably by using spectra of higher resolution and better signal-to-noise ratio.

%--------------------------------------------------------------------

\begin{acknowledgements}
{We thank the anonymous referee for useful comments that significantly helped to improve the paper. This study has benefited from the activities of the "ChETEC" COST Action (CA16117), supported by COST (European Cooperation in Science and Technology) and from the European Union’s Horizon 2020 research and innovation programme under grant agreement No 101008324 (ChETEC-INFRA). JK acknowledges support from European Social Fund (project No 09.3.3-LMT-K-712-19-0172) under grant agreement with the Research Council of Lithuania (LMTLT). This work has made use of the VALD database, operated at Uppsala University, the Institute of Astronomy RAS in Moscow, and the University of Vienna.}
\end{acknowledgements}

\begin{appendix}

\section{Differences between Ba abundances determined in different observing samples}

\begin{table}
	\begin{center}
		\caption{Target stars common to different observing programs.\label{app-tab:common-stars}}
		\begin{tabular}{lll}
			\hline\hline
			\noalign{\smallskip}
			072.D-0777(A) & 073.D-0211(A) & 088.D-0026(A) \\
			\hline\noalign{\smallskip}
			B-1256        &  --           &   R287  \\
			F-1389        &   1389        &   --    \\
			B-3449        &  --           &   R563  \\
			F-3476        &  --           &   R317  \\
			F-4373        &   4373        &   --    \\
			--            &   5172        &   R259  \\
			B-5362        &  --           &   R253  \\
			F-7711        &   7711        &   --    \\
			--            &   7904        &   R443  \\
			B-7993        &  --           &   R277  \\
			B-9163        &  --           &   R800  \\
			--            &   9518        &   R237  \\
			--            &   9717        &   R682  \\
			B-9997        &  --           &   R248  \\
			F-10198       &  --           &   R756  \\
			F-10527       &  --           &   R782  \\
			--            &  10994        &   R795  \\
			F-12408       &  --           &   R784  \\
			F-13668       &  13668        &   --    \\
			B-13795       &  13795        &   R752  \\
			B-13853       &  --           &   R246  \\
			B-14583       &  14583        &   --    \\
			F-15451       &  15451        &   --    \\
			--            &  15552        &   R381  \\
			--            &  16597        &   R778  \\
			B-16667       &  --           &   R790  \\
			B-17819       &  --           &   R245  \\
			--            &  21369        &   R249  \\
			F-23236       &  23236        &   --    \\
			F-24463       &  24463        &   --    \\
			B-29146       &  --           &   R256  \\
			--            &  29490        &   R231  \\
			F-30104       &  30104        &   --    \\
			B-30463       &  30463        &   --    \\
			B-30949       &  --           &   R766  \\
			B-32730       &  32730        &   --    \\
			B-35878       &  35878        &   --    \\
			B-38976       &  --           &   R762  \\
			B-41429       &  --           &   R392  \\
			B-42866       &  42866        &   R656  \\
			B-42887       &  --           &   R760  \\
			F-43632       &  43632        &   R512  \\
			B-43852       &  43852        &   R704  \\
			B-43889       &  43889        &   R450  \\
			\hline
		\end{tabular}
	\end{center}
\end{table}

\begin{figure}[tb]
	\resizebox{\hsize}{!}{\includegraphics{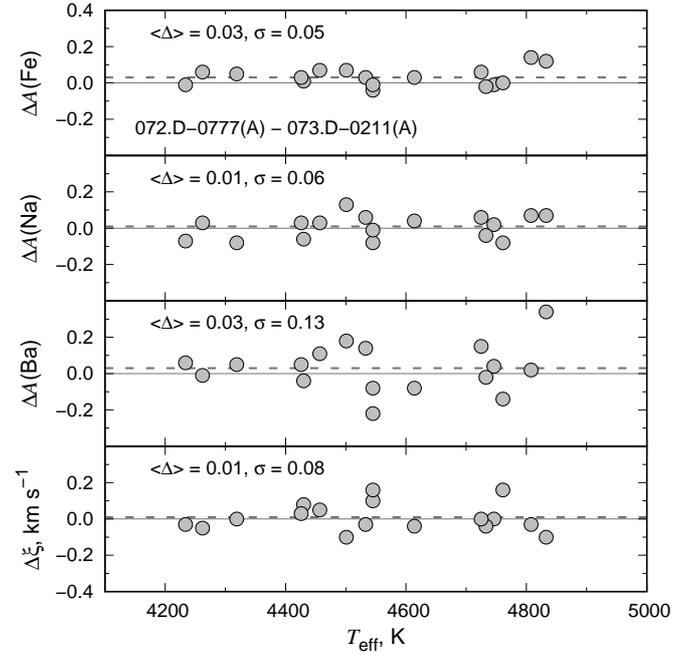}}
	\caption{Comparison of the abundances (top three pannels) and microturbulence velocity values between the common stars in the programs 072.D-0777(A) and 073.D-0211(A). Grey dashed line shows average value of the difference.\label{app-fig:compar-abund1}}
	\label{fig:fra-car-common}
\end{figure}

\begin{figure}[tb]
	\resizebox{\hsize}{!}{\includegraphics{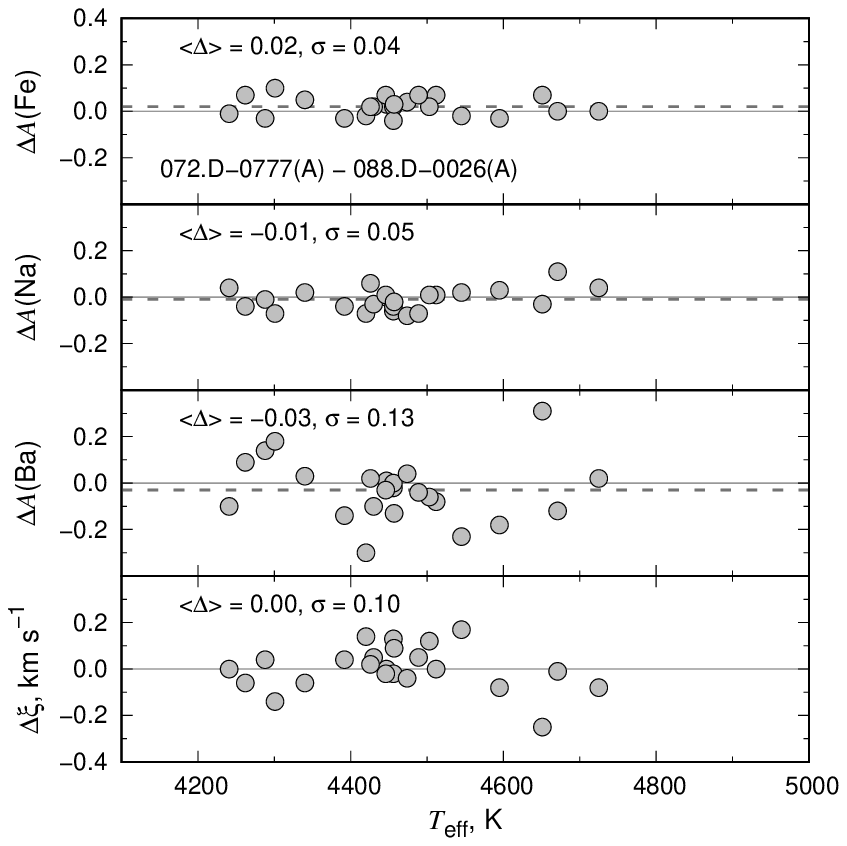}}
	\caption{Comparison of the abundances (top three pannels) and microturbulence velocity values between the common stars in the programs 072.D-0777(A) and 088.D-0026(A). Grey dashed line shows average value of the difference.\label{app-fig:compar-abund2}}
	\label{fig:fra-mcd-common}
\end{figure}

\begin{figure}[tb]
	\resizebox{\hsize}{!}{\includegraphics{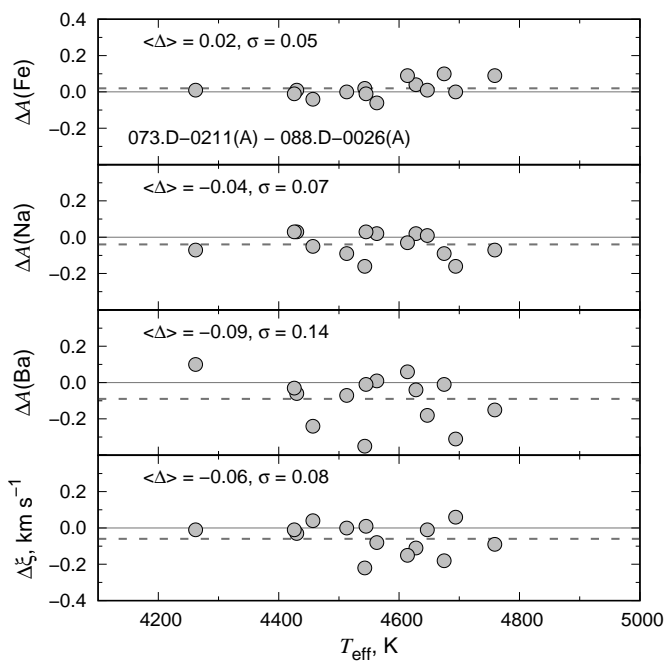}}
	\caption{Comparison of the abundances (top three pannels) and microturbulence velocity values between the common stars in the programs 073.D-0211(A) and 088.D-0026(A). Grey dashed line shows average value of the difference.\label{app-fig:compar-abund3}}
	\label{fig:car-mcd-common}
\end{figure}

As was indicated in Sect.~\ref{sect:abnd-analysis}, a number of target RGB stars have been observed in several observing programs (Table~\ref{app-tab:common-stars}). A comparison of Fe, Na, and Ba abundances obtained using spectra that were acquired in the different programs is provided in Fig.~\ref{app-fig:compar-abund1}--\ref{app-fig:compar-abund3}. In most cases, differences between the average abundances obtained in various samples was very small and did not exceed 0.03\,dex (Figs~\ref{fig:fra-car-common}-\ref{fig:fra-mcd-common}).

The final abundances of the target RGB stars were further corrected for these small systematic shifts. Since the program 072.D-0777(A) had the largest number of stars, we chose to apply abundance shifts relative to this sample. For stars in 073.D-0211(A) sample we applied abundance shifts of +0.03, +0.01, +0.03\,dex for the abundances of Fe, Na, and Ba, respectively. In case of the sample 088.D-0026(A), abundances of Fe, Na, and Ba were corrected by +0.02, $-0.01$, and $-0.03$\,dex, respectively. The final abundance ratios (Table~\ref{app-tab:abund-indiv-stars}) were computed using Solar reference values derived in Sect.~\ref{subsect:ref-abnd}.

\section{The \ion{Fe}{i} and \ion{Fe}{ii} lines used in the abundance analysis}

The list of \ion{Fe}{i} and \ion{Fe}{ii} lines and their atomic parameters is provided in the Table~\ref{app-tab:iron-list}. We note that abundances from \ion{Fe}{ii} lines were only determined with the purpose of checking the agreement between the surface gravities obtained from photometry and those derived using \ion{Fe}{ii} lines; however, they were not used in further analysis. Thus, Fe abundances used throughout this paper are those that were obtained using \ion{Fe}{i} lines.

\begin{table}
	\begin{center}
		\caption{The list of iron spectral lines used in the abundance determination.
			\label{app-tab:iron-list}}
		\begin{tabular}{cccc}
			\hline\hline
			\noalign{\smallskip}
			$\lambda$, nm & $\chi$, eV & log \textit{gf} & Ion. stage\\
			\hline\noalign{\smallskip}
			612.79070 &  4.140 & -1.398 & \ion{Fe}{i} \\
			615.16180 &  2.180 & -3.300 & \ion{Fe}{i} \\
			616.53600 &  4.140 & -1.460 & \ion{Fe}{i} \\
			617.33360 &  2.220 & -2.810 & \ion{Fe}{i} \\
			618.02040 &  2.730 & -2.650 & \ion{Fe}{i} \\
			618.79900 &  3.940 & -1.580 & \ion{Fe}{i} \\
			620.03130 &  2.610 & -2.310 & \ion{Fe}{i} \\
			621.34300 &  2.220 & -2.550 & \ion{Fe}{i} \\
			621.92810 &  2.200 & -2.410 & \ion{Fe}{i} \\
			622.92283 &  2.845 & -2.805 & \ion{Fe}{i} \\
			623.26410 &  3.650 & -1.130 & \ion{Fe}{i} \\
			624.63188 &  3.602 & -0.733 & \ion{Fe}{i} \\
			625.25554 &  2.404 & -1.687 & \ion{Fe}{i} \\
			626.51340 &  2.180 & -2.510 & \ion{Fe}{i} \\
			627.02250 &  2.860 & -2.570 & \ion{Fe}{i} \\
			627.12788 &  3.332 & -2.703 & \ion{Fe}{i} \\
			630.15012 &  3.654 & -0.718 & \ion{Fe}{i} \\
			631.58115 &  4.076 & -1.710 & \ion{Fe}{i} \\
			632.26860 &  2.590 & -2.280 & \ion{Fe}{i} \\
			633.53308 &  2.198 & -2.177 & \ion{Fe}{i} \\
			633.68243 &  3.686 & -0.856 & \ion{Fe}{i} \\
			634.41490 &  2.430 & -2.890 & \ion{Fe}{i} \\
			638.07430 &  4.190 & -1.270 & \ion{Fe}{i} \\
			660.91100 &  2.560 & -2.610 & \ion{Fe}{i} \\
			670.35670 &  2.760 & -3.010 & \ion{Fe}{i} \\
			672.66660 &  4.610 & -1.010 & \ion{Fe}{i} \\
			675.01530 &  2.420 & -2.580 & \ion{Fe}{i} \\
			680.68450 &  2.730 & -3.090 & \ion{Fe}{i} \\
			681.02630 &  4.610 & -0.940 & \ion{Fe}{i} \\
			684.36560 &  4.550 & -0.780 & \ion{Fe}{i} \\
			685.51620 &  4.560 & -0.570 & \ion{Fe}{i} \\
			685.81500 &  4.610 & -0.910 & \ion{Fe}{i} \\
			691.66820 &  4.150 & -1.260 & \ion{Fe}{i} \\
			614.92580 &  3.890 & -2.720 & \ion{Fe}{ii} \\
			623.83920 &  3.870 & -2.520 & \ion{Fe}{ii} \\
			624.75570 &  3.890 & -2.320 & \ion{Fe}{ii} \\
			636.94620 &  2.891 & -4.160 & \ion{Fe}{ii} \\
			\hline
		\end{tabular}
	\end{center}
\end{table}

\section{Abundances of Ba in the target RGB stars with $4600<\teff<4700$\,K\label{app-sect:ba-abund-teff46-47}}

\begin{figure}[tb]
\resizebox{\hsize}{!}{\includegraphics{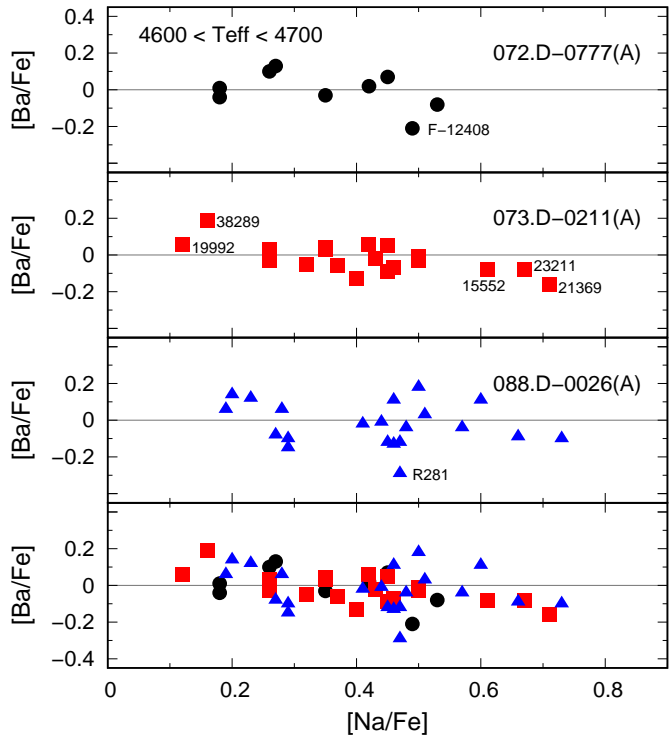}}
\caption{Barium and sodium abundances in the effective temperature range $4600 < \teff < 4700$\,K. Labels next to the outlying data points show identifications of the stars.\label{app-fig:ba-abund-teff-bin4600-4700}}
\label{fig:abn-t46-t47}
\end{figure}

As noted in Sect.~\ref{sect:results}, our data hint to a weak Ba--Na anti-correlation. Further analysis has revealed that this may be an artefact caused by several spurious outliers with $4600<\teff<4700$\,K that were observed in one of the programs, 073.D-0211(A) (Fig.~\ref{app-fig:ba-abund-teff-bin4600-4700}). No such anti-correlation is seen in any other $\Delta\teff=100$\,K wide effective temperature bins. Indeed, no statistically significant anti-correlation is left in the $4600<\teff<4700$\,K either when stars 19992, 38289, and 21369 are excluded. We therefore conclude that there is no statistically significant Ba--Na correlation in the analysed sample of 261 RGB stars in 47~Tuc.

\section{Abundances of Fe, Na, and Ba in 261 target RGB stars in 47~Tuc}

Abundances of Fe (LTE), Na (NLTE), and Ba (NLTE) obtained in 261 RGB stars in 47~Tuc are provided in Table~\ref{app-tab:abund-indiv-stars}.

\clearpage

\setlength{\tabcolsep}{5pt}

\begin{table*}
\begin{center}
\caption{Abundances of Fe, Na, and Ba determined in the sample of 261 stars in 47~Tuc. In cases when abundances were determined using data from several observing programs, microturbulence velocities and abundances obtained using spectra from each individual program are provided. The IDs of individual stars and IDs of the corresponding observing programs are given in the last two columns, respectively (072.D-0777(A), PI: P.~Fran\c{c}ois; 073.D-0211(A), PI: E.~Carretta; 088.D-0026(A), PI: I.~McDonald). The complete table is available in electronic form. \label{app-tab:abund-indiv-stars}}
\resizebox{\hsize}{!}{
\begin{tabular}{cccccccccccc}
\hline\hline
\noalign{\smallskip}
GAIA Source ID  & $\teff$ & $\log g$  & $\xi_{\rm micro}$ &  $A{\rm (Fe)}$  &  [Fe/H] &  $A{\rm (Na)}$  &  [Na/Fe]  &  $A{\rm (Ba)}$  &  [Ba/Fe] & ID & Obs. program  \\
&  K     &           &      \,km\,s$^{-1}$         &         &         &         &           &         &       &    \\
\hline\noalign{\smallskip}
4689575185928019840	&	4275	&	1.45	&	1.47	&	6.77	&	-0.78	&	5.58	&	0.19	&	1.54	&	0.15	&	R761	&	088.D-0026(A)	\\

4689575289007212416	&	4231	&	1.34	&	1.56	&	6.77	&	-0.78	&	5.60	&	0.21	&	1.54	&	0.15	&	R772	&	088.D-0026(A)	\\

4689581263316313856	&	4553	&	2.07	&	1.53	&	6.77	&	-0.78	&	5.63	&	0.24	&	1.39	&	0.00	&	R786	&	088.D-0026(A)	\\

4689613840634674048	&	4271	&	1.48	&	1.48	&	6.82	&	-0.73	&	5.66	&	0.22	&	1.38	&	-0.06	&	5277	&	073.D-0211(A)	\\

4689614837066802944	&	4630	&	2.09	&	1.40	&	6.79	&	-0.76	&	5.84	&	0.43	&	1.28	&	-0.13	&	6808	&	073.D-0211(A)	\\

4689614970198135296	&	4245	&	1.34	&	1.57	&	6.78	&	-0.77	&	5.56	&	0.16	&	1.36	&	-0.04	&	B-13396	&	072.D-0777(A)	\\

4689618238680828800	&	4303	&	1.50	&	1.58	&	6.75	&	-0.80	&	5.63	&	0.26	&	1.33	&	-0.04	&	B-9254	&	072.D-0777(A)	\\

4689618891515852160	&	4704	&	2.19	&	1.44	&	6.81	&	-0.74	&	6.10	&	0.67	&	1.42	&	-0.01	&	R198	&	088.D-0026(A)	\\

4689619166393741824	&	4512	&	1.96	&	1.46	&	6.82	&	-0.73	&	5.73	&	0.29	&	1.36	&	-0.08	&	R199	&	088.D-0026(A)	\\

4689621709015302400	&	4560	&	2.04	&	1.29	&	6.81	&	-0.74	&	5.88	&	0.45	&	1.42	&	-0.01	&	R677	&	088.D-0026(A)	\\
\ldots	&	\ldots	&	\ldots	&	\ldots	&	\ldots	&	\ldots	&	\ldots	&	\ldots	&	\ldots	&	\ldots	&	\ldots	&	\ldots	\\
\hline
\end{tabular}
}
\end{center}
\end{table*}

\end{appendix}

\end{document}